\documentclass[letterpaper,12pt,final]{iopart}
\pdfoutput=1
\usepackage{graphicx}
\usepackage{subfigure}
\usepackage{wrapfig}

\usepackage[colorlinks=blue]{hyperref}  
\hypersetup{pdftitle={Studying the effect of Polarisation in Compton scattering in the undergraduate laboratory}}

\usepackage[usenames, dvipsnames]{color}

\usepackage[normalem]{ulem}
\begin{document}

\title{Studying the effect of Polarisation in Compton scattering in the undergraduate laboratory}

\author{P.~Knights$^{1}$, F.~Ryburn$^{2}$\footnote[1]{Ogden Trust Summer Intern at the University of Birmingham.}, G.~Tungate$^{1}$, K.~Nikolopoulos$^{1}$}
\address{$^{1}$ School of Physics and Astronomy, University of Birmingham, B15 2TT, United Kingdom}
\address{$^{2}$ School of Physics and Astronomy, University of Oxford, OX1 3RH, United Kingdom}
\ead{k.nikolopoulos@bham.ac.uk}

\begin{abstract}
An experiment for the advanced undergraduate laboratory allowing
students to directly observe the effect of photon polarisation on
Compton scattering is described. An initially unpolarised beam of
photons is polarised via Compton scattering and analysed through a
subsequent scattering. The experiment is designed to use equipment
typically available at an undergraduate physics laboratory. The
experimental results are compared with a Geant4 simulation and
geometry effects are discussed.
\end{abstract}

%
%
\submitto{\EJP}
%
%

\section{Introduction} 
A large selection of experiments in the domain of nuclear and particle
physics is available in undergraduate laboratories. A typical example,
is the study of the Compton effect~\cite{Compton:1923zz} in $\gamma$
rays that found its place in the undergraduate training programme in
the 1960s, and is still an important part of the training of young
physicists. Various forms of the experiment have been proposed for the
undergraduate laboratory, from the measurement of $\gamma$-ray
absorption coefficients for probing the characteristics of the
scattered photons~\cite{Bartlett1964a}, to experiments involving
precision spectroscopy and timing~\cite{Bartlett1964d, French1965,
  Stamatelatos1972}. The increasing precision that can be achieved
with inexpensive means has led to extensions of laboratory exercises
which demonstrate the relativistic energy-momentum relation for
electrons~\cite{Jolivette1994} and the precision of
detectors~\cite{Hieronymus1998}.

Despite the variety of experiments available, students rarely observe
polarisation effects in the energy regime relevant for nuclear and
particle physics. Such polarisation effects can become readily
observable through Compton scattering.
An experiment is described where a beam of initially unpolarised
photons undergoes two subsequent Compton scatterings: The first
results in the polarisation of the beam, while the second analyses the
degree of polarisation. An asymmetry in the counting rate is observed
after the second scattering in the planes parallel and perpendicular
to the first scattering plane.
This arrangement has been investigated theoretically in
Ref.~\cite{PhysRev.74.1813}, where the differential cross section for
initially unpolarised photons undergoing two scatterings before being
detected is derived.

Similar experiments have been performed earlier with hard
X-rays~\cite{PhysRev.50.875}, as well as with $\gamma$ rays from
$^{60}$Co decays~\cite{PhysRev.85.58,Hudson1968}, however here the
experiment is carried out successfully using sources typically
available in the advanced undergraduate laboratory.

\section{Photon polarisation in Compton effect}
\begin{figure}[h!]
\centering
\subfigure[\label{fig:polarisation}]{\includegraphics[width=0.35\textwidth]{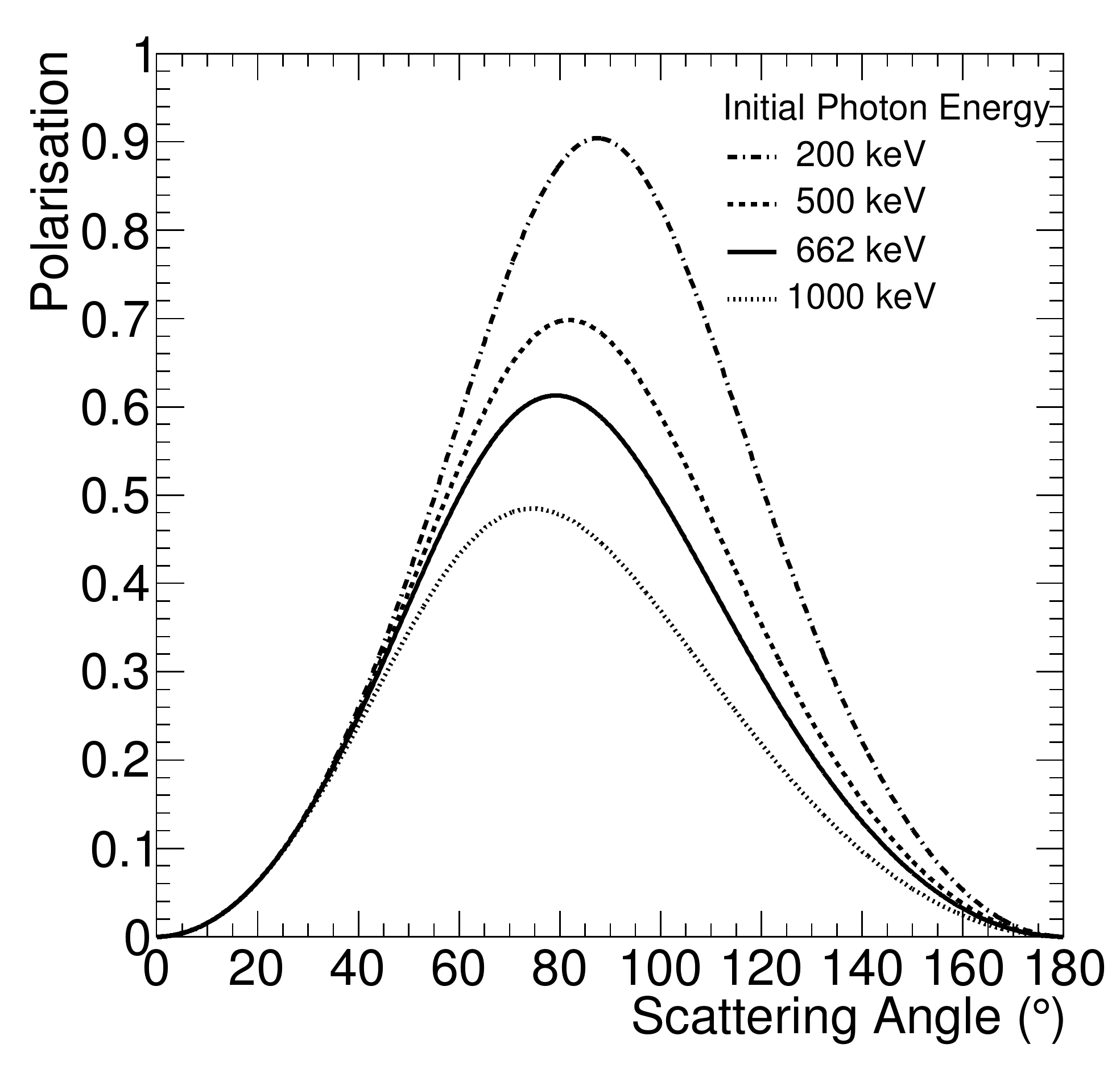}}
\subfigure[\label{fig:polarisationPA}]{\includegraphics[width=0.35\textwidth]{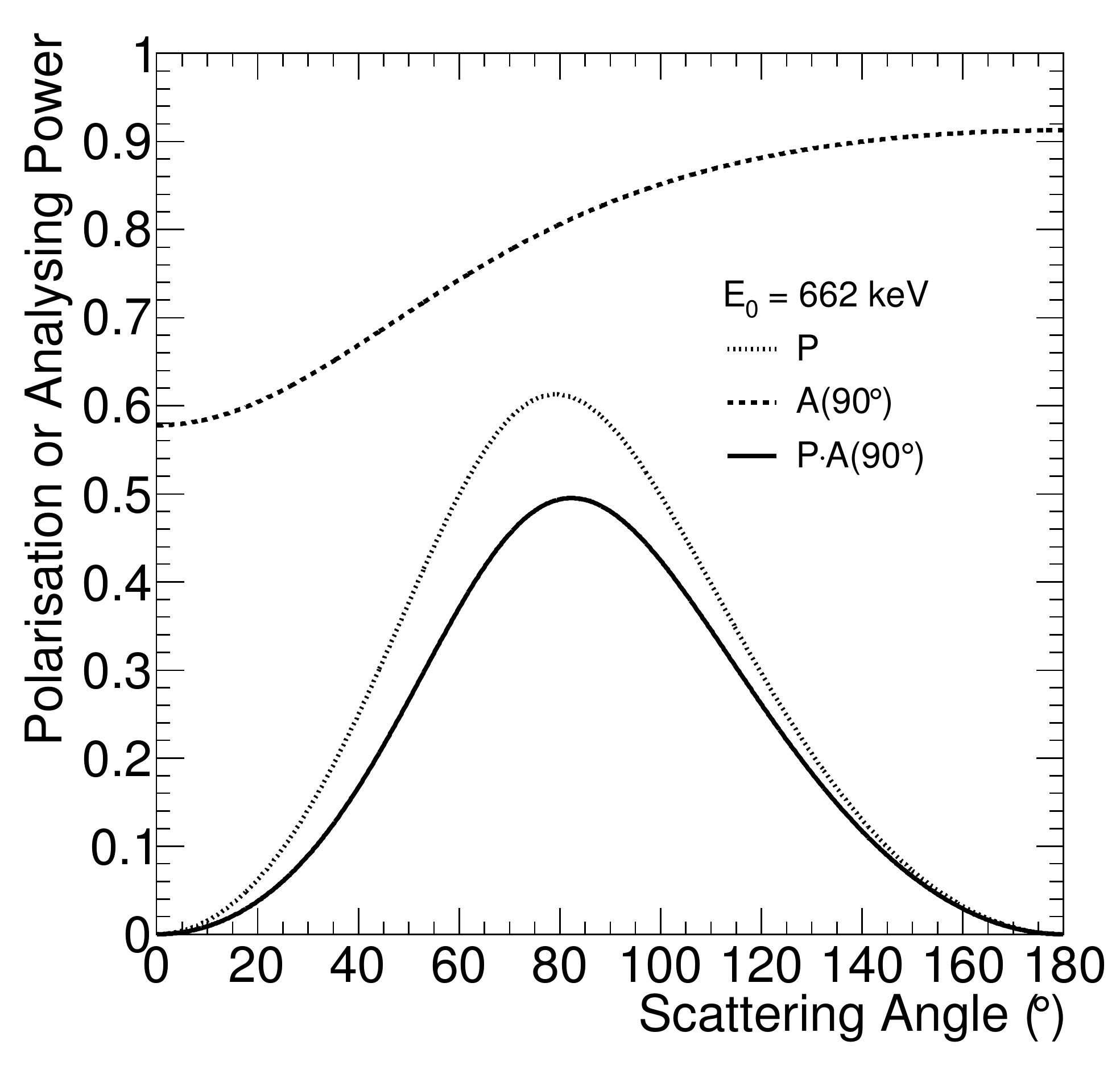}}
\caption{\subref{fig:polarisation} Polarisation of photons following Compton scattering, as a function of the scattering angle.
For photons of 662 keV a polarisation of 61.3\% is obtained for a scattering angle of 80$^{\circ}$.
\subref{fig:polarisationPA} Polarisation, analysing power at 90$^\circ$, and measurable polarisation $PA(90^\circ)$, for an initially inpolarised beam of photons with energy 662 keV scattered at various angles. The maximum measurable polarisation is 49.5\% at 82.3$^\circ$.
\label{fig:polarisationAll}}
\end{figure}

The differential cross-section for Compton scattering is given by the
Klein-Nishina formula~\cite{Heitler1954}:
\begin{equation}
\label{eq:Klein-Nishina}
\frac{d\sigma}{d\Omega} = \frac{r_0^2}{4}\left(\frac{E_1}{E_0}\right)^2 \left(\frac{E_0}{E_1} +\frac{E_1}{E_0} -2 + 4\cos^2\Theta \right),
\end{equation}
where $r_0^2$ is the classical electron radius, $E_0$ is the energy of
the incident photon, $E_1$ is the energy of the scattered photon, and
$\Theta$ is the angle between the photon polarization vectors before
and after the scattering.  For polarised photons this results in the
photon angular distribution after scattering not being symmetric
around the initial photon momentum, a property that is extensively
used in $\gamma$-ray polarimeters.
For an initially unpolarised beam of photons, the scattered photons
will be partially polarised. The degree of polarisation
\begin{equation}
P=\frac{I_\perp-I_\parallel}{I_\perp+I_\parallel}= \frac{\sin^2\theta}{\frac{E_0}{E_1}+\frac{E_1}{E_0} - \sin^2\theta},
\end{equation}
where $I$ is the photon intensity and $\perp$($\parallel$) denote a photon polarised perpendicular
(parallel) to the plane of the scattering. The polarisation depends on the energy of
the incident photon and the scattering angle as shown in
Figure~\ref{fig:polarisation}. 

The analysing power A is defined as 
$A=\frac{{\cal N}_\parallel-{\cal N}_\perp}{{\cal N}_\parallel+{\cal N}_\perp}$, 
where ${\cal N}_\parallel$ (${\cal N}_\perp$) denotes the count rate
detected for photons polarised parallel (perpendicular) to the normal
of the scattering plane. Using Eq.~\ref{eq:Klein-Nishina} the analysing power is obtained:
\begin{equation}
A = \frac{\sin^2\theta'}{\frac{E_1}{E_2}+\frac{E_2}{E_1}-\sin^2\theta'}~,
\end{equation}
where $E_2$ is the energy of the photon following the second scattering at angle $\theta'$. 
Should photons be polarized by an initial scattering, the degree of polarization can be measured in a second scattering by measuring the ratio
\begin{equation}
R=PA=\frac{{N}_\parallel-{N}_\perp}{{N}_\parallel+{N}_\perp},
\end{equation}
where ${N}_\parallel$ is the count rate for coplanar scattering and
${N}_\perp$ is the count rate when the two scattering planes are at
right angles to each other. In Figure~\ref{fig:polarisationPA} the calculated
polarisation $P$ for 662~keV photons as a function of the scattering
angle is presented, along with the analysing power $A(90^\circ)$ for
the scattered photons undergoing a second scattering of
$90^\circ$. The measurable asymmetry, $PA(90^\circ)$, can be seen to
peak at approximately $80^\circ$ with a value of approximately
0.49. The function is slowly varying with angle which enables a
reasonable estimate of the effective asymetry to be made provided the
solid angles are small and known.

Thus, the energy of the initial gamma ray should be low to ensure a
high degree of polarisation and a large analysing power. However, if
the energy is too low, the photoelectric effect will dominate over
Compton scattering.  As a result, the 662~keV gamma ray from
Caesium-137 is ideally suited for this experiment.

\section{Experimental Arrangement}
\begin{figure}[h!]
\centering
\subfigure[\label{fig:setupSketch}]{\includegraphics[width=0.35\textwidth]{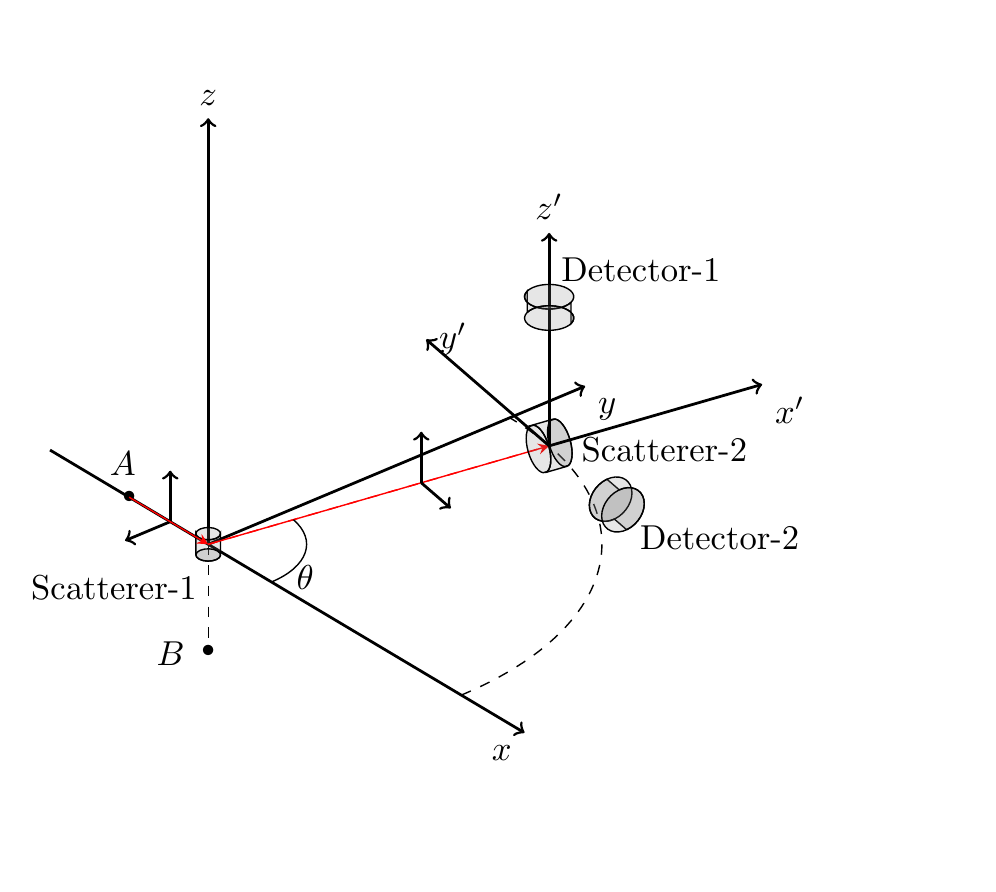}}
\subfigure[\label{fig:setup}]{\includegraphics[width=0.35\textwidth]{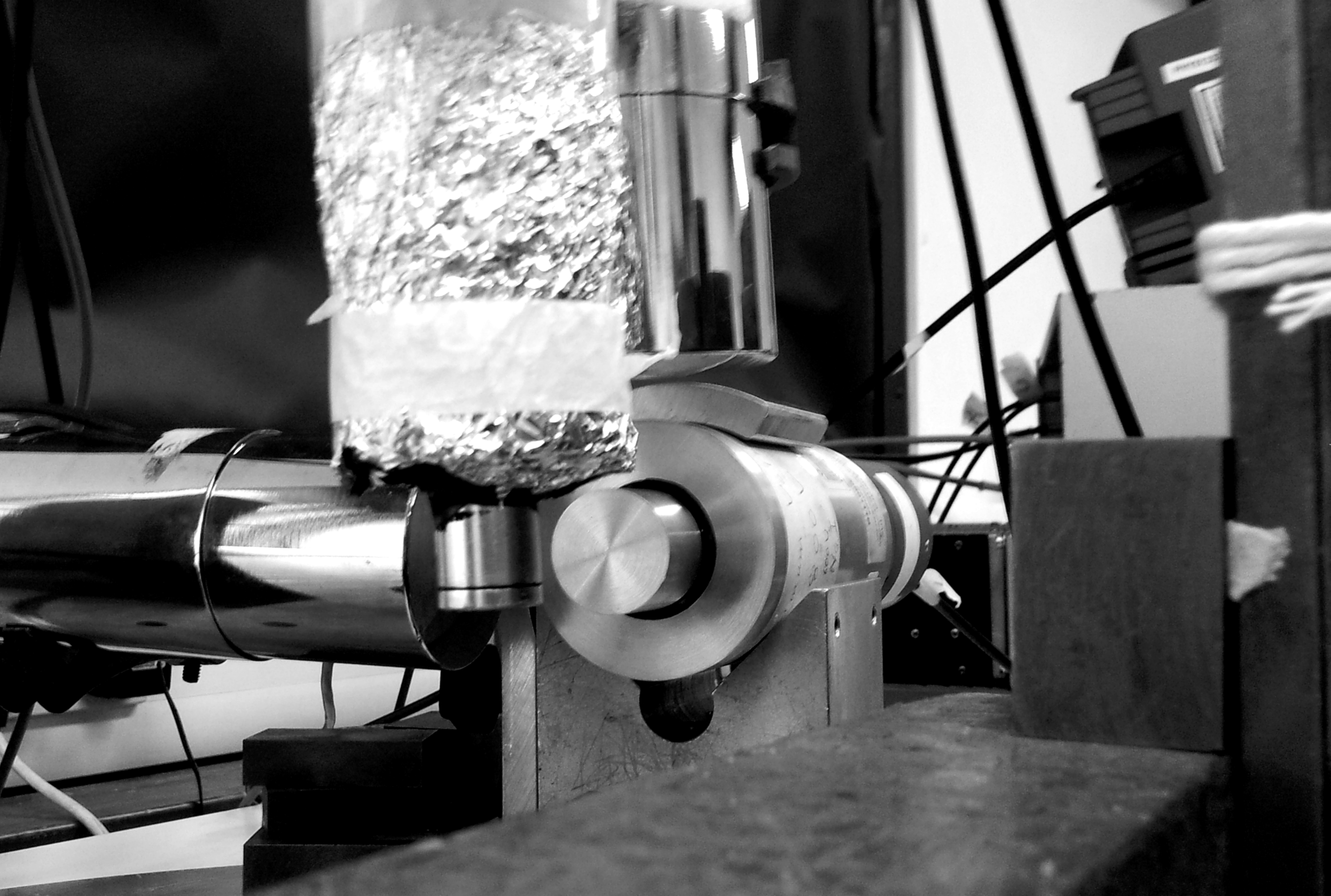}}
\caption{\subref{fig:setupSketch} Layout of the experimental set-up. 
\subref{fig:setup} Photograph of the experimental set-up. }
\end{figure}

The experimental layout is shown in Figure~\ref{fig:setupSketch},
while a general view of the experiment is given in
Figure~\ref{fig:setup}. The apparatus consists of four NaI(Tl)
scintillators, presented in Table~\ref{tab:detector}. For the main
experiment a collimated Caesium-137 source with an apperture of 3~mm
is used, nominally at position ``A''.  Caesium-137 undergoes $\beta^-$
decay to Barium-137, and emits a 661.7 keV photon from the
de-excitation of the daughter nucleus. The photon rate from the collimated source
was of the order of a few tens of thousands per second.

The photon, after leaving the source, impinges on Scatterer-1, at a
distnace of 13~cm, and emerges at an angle of $80^\circ$, with an
energy of 320 keV. Subsequently, it scatters off Scatterer-2, at a
distance of 7.8~cm, and emerges at $90^\circ$ with 197 keV, and is
detected in Detector-1 or Detector-2, which each lie at 7~cm from
Scatterer-2. Detector-1 is in the plane perpendicular to the plane of
the first scattering, while Detector-2 lies in the plane of the first
scattering. Appropriate shielding was added using blocks of lead. The
aim of the shielding was to obstruct direct line-of-sight from the
source to Scatterer-2 and the detectors, as well as from Scatterer-1
to the detectors.

As estimated from a Geant4 simulation of the experiment, see
Section~\ref{sec:simulation}, in the described experimental
arrangement $40\%$ of the photons leaving the source interact in
Scatterer-1, with $33\%$ of those interacting being fully
absorbed. Approximately $1$ in $1000$ photons from the source arrive at
Scatterer-2 having interacted with Scatterer-1, while $1$ in $100000$
photons from the source arrive at either Detector-1 or Detector-2. Of
the photons observed in Detector-1 and Detector-2, 
78.5\% have undergone a single scattering in Scatterer-1, while 74.3\% have scattered only once in Scatterer-2.

\begin{table}[h!]
\centering
\caption{The dimensions of the cylindrical NaI(Tl) scintillators used
  in the experiment, along with their measured energy
  resolution.\label{tab:detector}}
\begin{tabular}{cccccc}
\hline\hline
Name &  Diameter & Length &  Energy Resolution \\
     & (in.) & (in.) & FWHM at 511 keV \\
\hline	
Scatterer-1 & 0.75 & 1 & 8.3\% \\
Scatterer-2 & 1   & 1 & 7.7\% \\
 Detector-1 & 2   & 2 & 7.0\% \\
 Detector-2 & 2   & 2 & 6.9\% \\
\hline\hline
\end{tabular}
\end{table}

\section{Detector calibration and characterisation}
The detectors were calibrated using photons of known energies from a
variety of radioisotopes~\cite{sonzogni2007nndc}: $^{68}$Ge,
$^{241}$Am, and $^{108{\rm m}}$Ag.
Germanium-68 decays by electron capture to Gallium-68 which itself
decays to Zinc-68 via $\beta^{+}$ decay. The annihilation of the
positron releases two back-to-back 511~keV photons. Americium-241
disintegrates through $\alpha$ decay to Neptunium-237, which in turn
emitts a 59.5~keV photon. Silver-108m undergoes electron capture to
Palladium-108 and emitts, among others, a 433.9~keV $\gamma$-ray. In
particular, Scatterer-1 was calibrated also using the 661.7 keV line
of Caesium-137 to allow monitoring of the photo-absorption. Beyond
energy calibration, these data have also been used to estimate the
resolution of the detectors, through the full-width at half-maximum
(FWHM) of the peaks in the obtained spectra.
The timing of the various detectors was studied and synchronized using
annihilation photons.

To avoid apparent asymmetries between the two detectors their
efficiency was measured and compared. Annihilation photons from
Germanium-68 were used in the coincidence technique: Detector-1 and
Detector-2 were placed symmetrically about the source, and the
detection of a photon in Detector-1 was used as the trigger signal to
search for the other photon in Detector-2, and vice-versa. The
efficiencies of the two detectors were found to be 28\% and 26\%
respectively. Based on this a correction is applied to the data and a
systematic uncertainty equal to the observed difference in
efficiencies is assigned to the result.

\section{Data acquisition}
\begin{figure}[h!]
\centering
\includegraphics[width=0.45\textwidth]{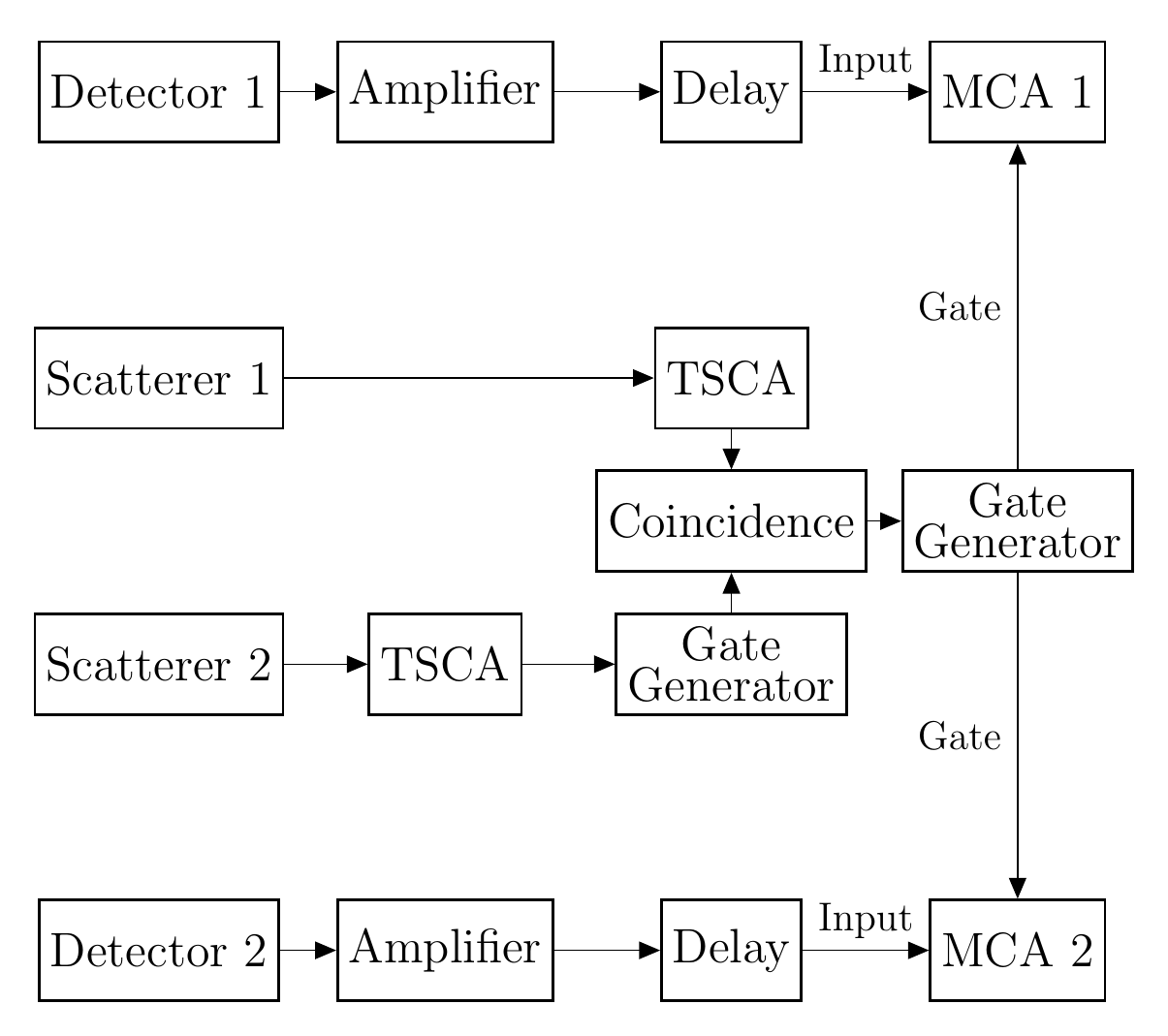}
\caption{Block diagram of the read-out electronics of the experimental arrangement.\label{fig:setupElectronics}}
\end{figure}

The pulses from Detector-1 and Detector-2 were amplified
and each fed into a multi-channel analyser (MCA). In order to record
signals which originated from scattering events, a logic signal was
devised using the pulses from Scatterer-1 and Scatterer-2 and used to
gate the two MCAs.
The anode signals from Scatterer-1 and Scatterer-2 were fed into two
timing single-channel analysers (TSCA)
to select the energies and discard events not originating from the
desired scattering pattern. A loose window was set for Scatterer-1,
approximately, from 80 to 500~keV. A window from, approximately,
90~keV to 153~keV was set for Scatterer-2. The outputs of the single
channel analysers were fed into a coincidence unit
to produce a logic pulse. The diagram of the circuit is shown in
Figure~\ref{fig:setupElectronics}.

The single channel analyser windows were set by splitting the signal
from the scatterer with one branch 
being fed to the timing single channel analyser as explained above, while the second branch was
fed to a pre-amplifier and subsequently to a spectroscopy amplifier.
The latter signal was appropriately delayed and fed to a multi-channel analyser
while the former was used as a gate. 

In the course of preparing the trigger signal, the background rates
were also studied. Without the presence of coincidence requirements
Detector-1 and Detector-2 recorded events at a rate of approximately $120\;{\rm s}^{-1}$.
By requiring a singal from Scatterer-1, these rates were reduced by a factor 20,
while requiring a signal from Scatterer-2 resulted in a reduction by a factor 1000, at which point about 50\% of the observed events were in the signal region.
If coincidence was required in both scatterers the detectors measured
a rate of $0.032\;{\rm s}^{-1}$ and $0.073\;{\rm s}^{-1}$, respectively, with about 80\% of the
events in the signal region.
To estimate the number of accidental coincidences, data were collected with the same
setup but with an additional delay between the coincidence signal
and the detector output. 

\section{Simulation of the Experiment}
\label{sec:simulation}
\begin{figure}[h!]
\centering
\includegraphics[width=0.60\textwidth]{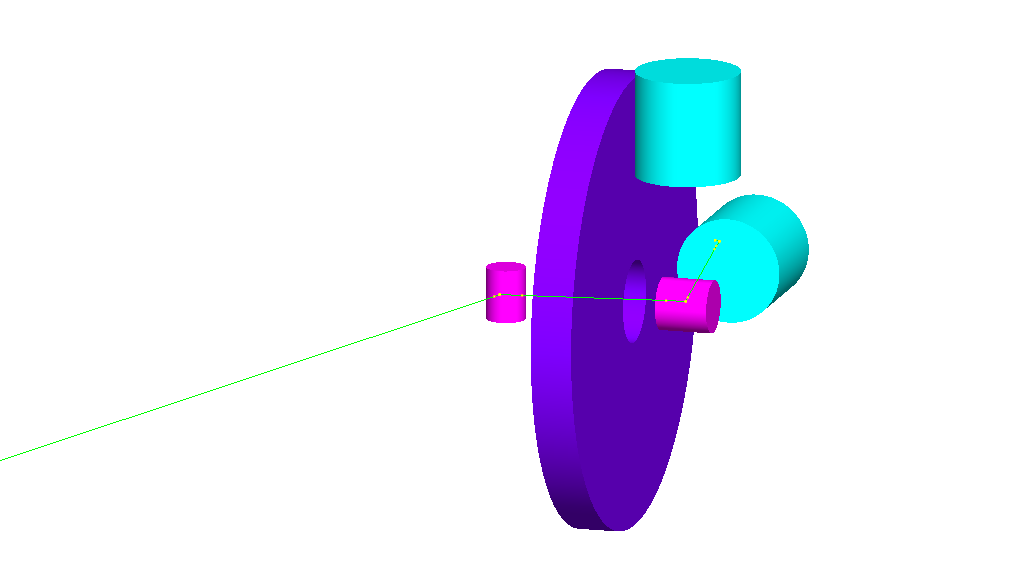}
\caption{Experimental layout in simulation. The scatterers are shown in purple, the detectors in teal, and the shielding in violet. The photon path is shown in green.\label{fig:simulationlayout}}
\end{figure}

The experimental arrangement has been replicated using the Geant4
simulation toolkit~\cite{Agostinelli:2002hh}, and the ratio $R$ has
been obtained using the same selection requirements as in the data
analysis. The experimental layout as implemented in simulation is shown in Figure~\ref{fig:simulationlayout}.

To highlight the effects of photon polarisation and to check for
apparent asymmetries of geometric origin, the simulation was also
performed ignoring polarisation effects using the
\texttt{G4EmLivermore} physics model. The estimated ratio was
$R_0=\left(4\pm 2\right)\%$. As a result, geometry effects are
found to induce a small asymmetry in the same direction as that
induced by polarisation. The uncertainty on $R_0$ includes a systematic
uncertainty due to potential mismatch between the energy thresholds
applied in simulation and the experiment.

A direct experimental estimate of the apparent asymmetry due to geometry effects could be obtained by positioning a Germanium-68 source between Scatterer-1 and Scatterer-2. Using the signal in the two scatterers, in coincidence, as a gate and counting the number of photons impinging on Detector-1 and Detector-2 having undergone Compton scattering in Scatterer-2, an estimate of $R_0$ is obtained. This estimate neglects the effect of Scatterer-1 in the overall apparent asymmetry of the experimental arrangement. 

\section{Results}
\begin{figure}[!b]
\centering
\includegraphics[width=0.49\textwidth] {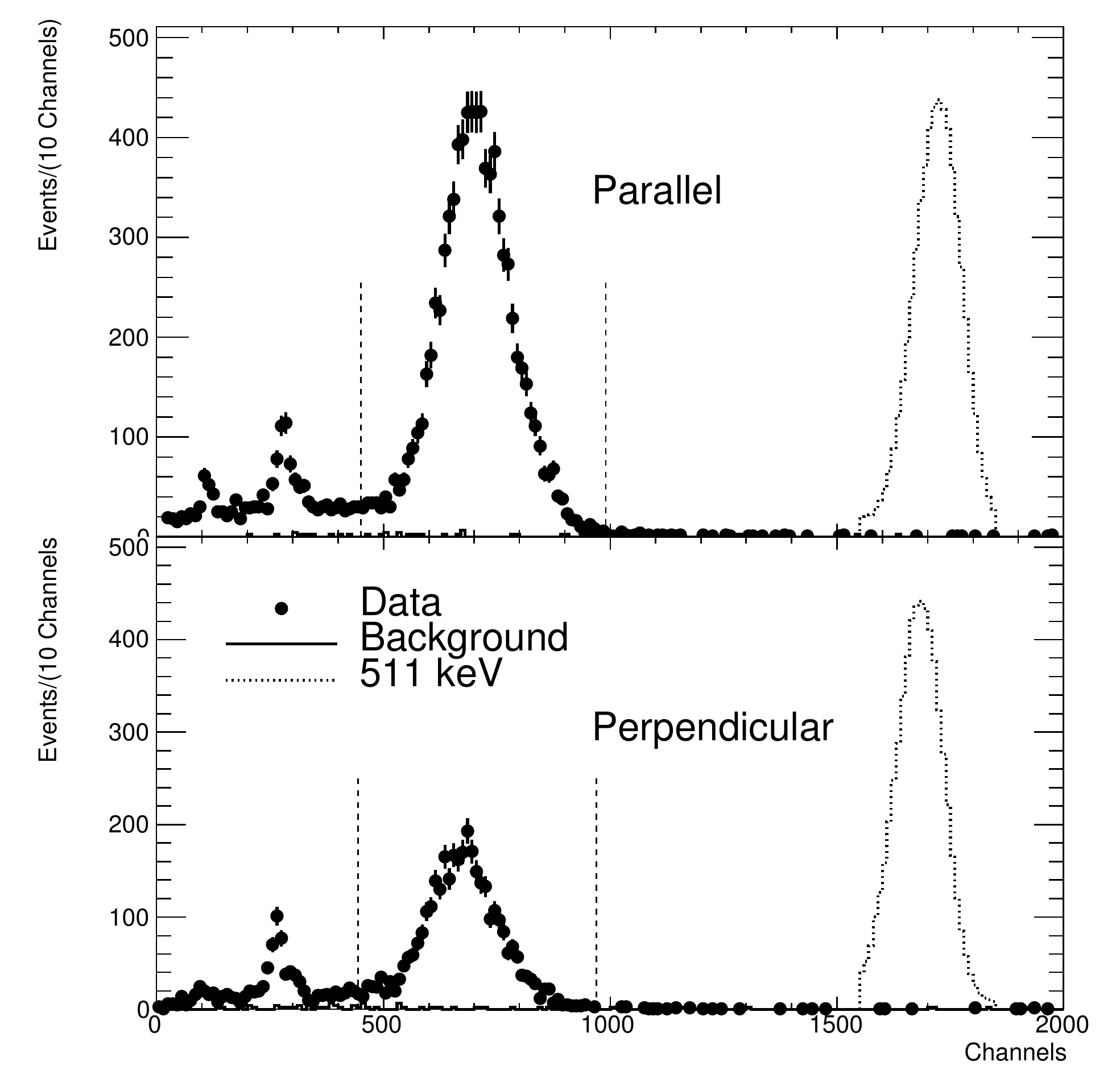}
\caption{The obtained experimental yields of events for the detector
  Parallel (above) and Perpendicular (below) to the first scattering
  plane. The continuous histogram denotes the contribution from random
  coincidences. The vertical dashed lines denote the signal region,
  between 135 and 295~keV.\label{fig:experimentalAsymmetry}}
\end{figure}

In Figure~\ref{fig:experimentalAsymmetry} the observed energy spectra
from the two detectors are shown after 39-hours of data-taking.
An asymmetry is readily observed. The peak at about 700 ADC channels
is the photo-peak from the absorption of the scattered photons
alongside the Compton continuum. At lower energies, the characteristic
X-rays of lead from the shielding and of iodine are also observed. A positron
annihilation photo-peak is overlaid to set the energy scale. From the
number of counts in the area denoted by the vertical dashed lines in
Figure~\ref{fig:experimentalAsymmetry}, $R=\left(45\pm 3\right)\%$ is
obtained, after the background from accidental coincidences is
subtracted. The quoted uncertainty includes the statistical
uncertainty of the measurement and the systematic uncertainty on the
uniformity of response of the two detectors. From this, the
polarisation of the photon beam after the first scattering is
estimated to be $P=\left(56\pm5\right)\%$, where a 5\% relative uncertainty on the analysing power has been included.

These estimates do not account for the finite geometry effects
discussed in the following section. If a correction is included for
the observed apparent asymmetry observed in the simulation when no
polarisation effects are included, the obtained ratio $R$ becomes
$\left(42\pm4\right)\%$ and the corresponding polarisation is
$P=\left(52\pm6\right)\%$, in agreement within uncertainties with the expectation.

As a cross-check, the experiment was repeated with a Caesium-137 source
placed in position ``B'', see Figure~\ref{fig:setupSketch}, effectively interchanging the roles of Detector-1 and Detector-2 as ``Parallel'' and ``Perpendicular''. An asymmetry was observed, albeit less pronounced and with larger uncertainties due to the inappropriate shape of the two scatterers for this orientation, which resulted in a wide spread of scattering angles and a larger geometrical asymmetry then the nominal experiment. 

\section{Discussion}
The experiment described allows for the observation of the effects of
polarisation in Compton scattering, starting from an initially
unpolarised source available in the undergradute physics laboratories.
Measurements were carried out over the period of hours, a time
interval appropriate for the undergraduate laboratory by allowing the
experiment to run overnight or a day.
It is noted that the choice of scattering angles probes the region
where polarisation effects play the most important role, but also
results to an experimental arrangement that is relatively insensitive
to small deviations of the detectors from the nominal positions.

The accidental coincidences, signals uncorrelated to scattering
events, in the experiment have been estimated by introducing an
arbitrary delay between the trigger signal, and the read-out of the
detectors. Nevertheless, in the obtained energy spectra contributions
are observed that seem to originate in-time with the scattering
events, but are inconsistent with the expected energy for the photons
following the desired scattering path.  These contributions are
consistent with the K-lines of Lead, 88~keV, and Iodine, 33~keV. Their
energy is low enough to not obscure the measurement of the fully
absorbed scattered photons.

Two different approaches have been implemented to control any
potential apparent asymmetry in the counting rates due to systematic
effects. In the first one, the photo-peak efficiency of the two
detectors is measured using a coincidence technique with annihilation
photons, to demonstrate the consistency. The measurement was performed
before and after data-taking. The second approach involves the
repetition of the experiment after moving the source from position
``A'' to position ``B'', effectively interchanging the role of the two
detectors as parallel and perpendicular to the plane of first
scattering. A possible extension of this measurement could involve the
measurement of the ratio $R$ as a function of the angular separation
of the planes of the second scattering. Furthermore, the experiment
may be performed with one detector, instead of two, by moving the same
detector between the two positions, ensuring the equality of response
of the detectors. However, in the interest of time it was deemed
useful to take data in parallel.

In the presented experimental arrangement both
scatterers were ``active'', that is they were read-out and 
contributed to the background rejection. Nevertheless, from the
background studies performed it is demonstrated that replacing the
first scatterer with a brass rod would still be a viable solution.

A few comments on the minimum instrumentation requirements for this
experiment are in line. In the experiment standard off-the-shelf 2"
and 1" NaI(Tl) detectors were used. As discussed earlier, one could
consider using the same detector for both Detector-1 and Detector-2 by
moving it between the two positions, at the expense of longer data
collecting periods. Nevertheless, in this case the required electronic
units are reduced: One less pre-amplifier, amplifier, high voltage
bias supply, and multi-channel analyser. The smallest detector,
Scatterer-1, was built in-house from a sealed NaI crystal and a
photo-multiplier tube (PMT). Sealed crystals and PMTs are readily
available from a number of sources. However, Scatterer-1 could be
still replaced by a standard 1" detector, or by an inactive scatterer,
such as a copper rod. In this latter case, the required electronic
units are also reduced: One less pre-amplifier, amplifier, high
voltage bias supply, single channel analyser channel, and no need for
a coincidence unit. Finally, a standard set of laboratory sources is
needed for the calibration of the detectors, while a strong
Caesium-137 source was required for the main polarization
measurement. For the final implementation in the teaching laboratory a
collimated $5\;{\rm MBq}$ source was used, together with the required
shielding.

\section{Summary}
An experiment suitable for the undergraduate physics laboratory that
demonstrates the effect of polarisation in Compton scattering was
presented. An initially unpolarised beam of photons that undergoes two
subsequent scatterings is employed. A coincidence technique, along
with appropriate shielding, suppresses adequately the background from
accidental coincidences.  Additional controls of the detectors'
efficiency and inversion of their position verify the observed effect
and provides the opportunity for a discussion of systematic
uncertainties and their role in measurements.

This exercise could provide an effective introduction to the role of
photon polarisation in nuclear and particle physics, allow for a
project-type laboratory course, and introduce the students in
background estimation techniques, systematic checks, and advanced
simulation techniques, usually not discussed in the undergraduate
laboratory.

\ack
This work has been performed at the Year 3 Nuclear Physics Laboratory
of the School of Physics and Astronomy at the University of
Birmingham, where this exercise is now offered to undergraduate
students. PK acknowledges support from the School of Physics and Astronomy during his
summer placement. FR acknowledges support from the Ogden Trust Summer Internship
program. Useful feedback on early versions of manuscript from Dr J.~Wilson, is
gratefully acknowledged.

\section*{References}
\bibliographystyle{ieeetr}
\bibliography{compton}
\end{document}